# Franck-Condon electron emission from polar semiconductor photocathodes

W. Andreas Schroeder[1], L. A. Angeloni[1], I-J. Shan[1], and L. B. Jones[2,3]

[1] Department of Physics, University of Illinois Chicago, Chicago, Il 60607-7059, USA
[2] ASTeC, STFC Daresbury Laboratory, Warrington, WA4 4AD Cheshire, United Kingdom
[3] Cockcroft Institute of Accelerator Science and Technology, Warrington, WA4 4AD Cheshire, United Kingdom

**ABSTRACT**. A presented analytical formulation of (optical)phonon-mediated and momentum-resonant Franck-Condon emission of photoexcited electrons from polar semiconductors is shown to be very consistent with (i) the observed emission properties of a Cesiated GaAs(001) photocathode at 808nm [J. Phys. D: Appl. Phys. **54**, 205301 (2021)] and (ii) the measured spectral emission properties of a GaN(0001) photocathode from just below its bandgap energy to 5eV. The theoretical analysis in the parabolic band approximation predicts the form of both the quantum efficiency and mean transverse energy of photoemission as a function of the photocathode's electron affinity and the electron temperature in the vicinity of its emission face. The good agreement between theory and experimental data also suggests that sub-10nm rms surface roughness effects are not significant for polar semiconductor photocathodes.

**Introduction**

Semiconductor materials, particularly polar semiconductors [1], are fast becoming the photocathode material of choice for front-end photo-injectors [2-7] employed in XFELs [8] and potentially UED systems [9-13]. In large part this is because they offer enhanced performance over traditional metal photocathodes at more accessible visible to near UV wavelengths (i.e. photon energies ħω less than 4eV) [14]. Quantum efficiencies (QEs) greater than 0.1% are routinely achieved for $Cs_2Te$ [5,15-17], $Cs_3Sb$ [18-22], and the bi-alkali photocathodes [2,23,24] due to their low electron affinity χ – the energy difference between the vacuum and the minimum of the semiconductor's conduction band into which electrons are photoexcited ($\chi = E_{vac.} - E_{CBM} = \phi - E_g$ , where ϕ is the photoelectric work function measured from the valence band maximum and $E_g$ is the semiconductor band gap). Single-crystal GaAs [25-28] and GaN (both wurtzite and zinc blende phases [29]), and III-V semiconductor strained superlattice heterostructures [30] have also been shown to allow spin-polarized electron beam generation – critical for studies of parity violation [31] and nucleon spin structure [32] and valuable for investigating magnetic properties and spin states in materials physics [33-35]. In these cases, Cesiation of the photocathode surface can generate a negative electron affinity (NEA) that leads to QEs well over 1% [36-44].

In contrast, measurements of the mean transverse energy (MTE) of the photoemitted electron distribution, defined using the rms transverse momentum ($\Delta p_T$) of the emitted electrons in the vacuum, $MTE = (\Delta p_T)^2/(2m_0)$, where $m_0$ is the free electron mass, have not generally conformed to theoretical expectations for semiconductor photocathodes. Although in many cases the observed MTE approaches the thermal limit close to threshold [20,24,44], standard photoemission models [45-47] and Monte Carlo based simulations [48,49] of photoexcited electron emission (i.e., transport dynamics and emission physics with transverse momentum conservation [50]) predict lower values of the MTE, especially for GaAs [48,51], due to expected emission from a low effective mass conduction band [52-55]. As a result, surface roughness [56,57] (or scattering) effects are often invoked to explain the discrepancy [48,57-59], or alternatively an undefined mechanism that ensures electrons photoexcited into the conduction band(s) of the photocathode attain the free electron mass prior to emission [45,51].

In this Letter, we propose an alternative explanation for photo-excited electron emission from polar semiconductor photocathodes that is based on an (optical)phonon-mediated Franck-Condon (FC) scattering mechanism [59,60]. This momentum-resonant emission process has been observed directly in prior work by Rameau *et al.* [61] on a Hydrogen-terminated diamond(001) photocathode – a material with known strong optical deformation potential scattering [62]. Phonon scattering has also been invoked to explain inelastic electron emission [63] and the emitted electron energy distribution curves [64] from NEA GaAs and the emission properties of PbTe [65]. Our developed theoretical description of this emission process, based on the parabolic band approximation (as outlined in the Appendix), is shown to be in very good agreement with experimental measurements performed on Cesiated *p*-type GaAs photocathodes for incident light at 808nm [44] and spectral characterization measurements performed on a $p$-type GaN(0001) photocathode. These two photocathode studies represent the long and short electron transport length limits since, for incident photon energies close to but greater than their fundamental band gap, the absorption depth in GaAs [66] is an order of magnitude greater than that in GaN [67,68]. More specifically, the product of the absorption coefficient $\alpha$ and the electron drift velocity $v_d$ towards the emitting photocathode face (due to internal and applied fields) is less than the optical phonon decay rate in in GaAs, but greater in GaN, implying that the photoexcited electron distribution returns to the lattice temperature before emission in GaAs, but will need to equipartition its energy with the optical phonon modes in GaN [69].

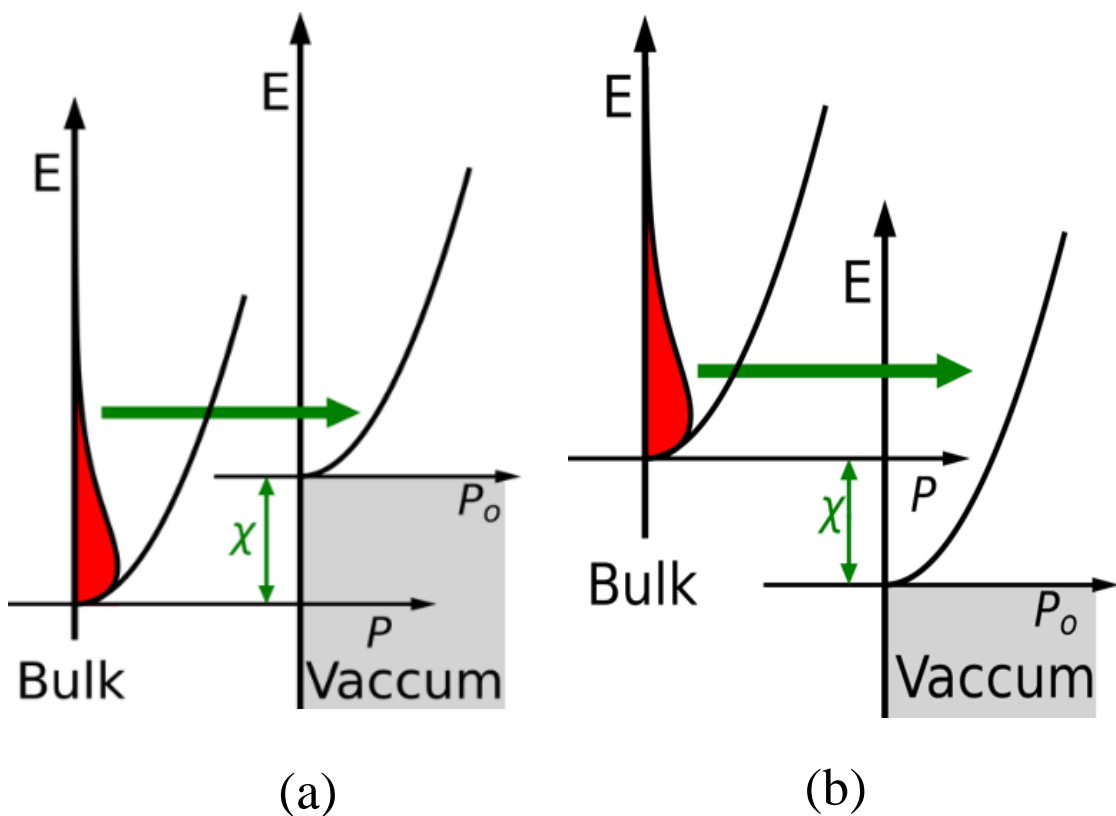

**Figure 1**
Schematic of Franck-Condon electron emission from a thermalized electron distribution in the parabolic band approximation for (a) positive electron affinity ($\chi > 0$) and (b) negative electron affinity ($\chi < 0$). A phonon of momentum **q** facilitates a momentum-resonant electron transition (green arrow) from the bulk band states $E(p)$ to the vacuum states $E(p_0)$.

**Franck-Condon Electron Emission**

Our theoretical interpretation of momentum-resonant FC electron emission begins with the electron flux $\mathbf{j}_z$ transmitted through the photocathode face into the vacuum. For a step work function potential barrier, it is straightforward to show that $j_z = p_{z0}/m_0$, under momentum resonant transmission over the barrier for which the barrier transmission probability $|T|^2 = 1$ (see the Appendix). As the transmitted vacuum electron momentum $p_{z0}$ perpendicular to the photocathode face (the longitudinal $z$ direction) can be written as $p_0\cos\theta$, where $\theta$ is the polar angle, we obtain an angular emission dependence similar to that expected from surface roughness effects [57], and indeed from free-electron-like metal photocathodes [46]. The remaining analysis presented in the Appendix deals with the mapping of the photoexcited electron distribution in the photocathode band structure onto the vacuum states under the momentum resonant assumption. For simplicity, only one isotropic parabolic emission band in the photocathode is considered for the cases of both positive and negative electron affinity $\chi$ (as shown in Figure 1) since this allows for an analytical solution. The resulting expressions for the QE and MTE associated with FC emission from a thermalized electron distribution at temperature $T_e$ are

$$QE = A(k_BT_e)^2[2k_BT_e + \chi]\, exp\left[-\frac{\chi}{k_BT_e}\right] \qquad 1(a)$$

$$MTE = k_BT_e\left[\frac{3k_BT_e+\chi}{2k_BT_e+\chi}\right] \qquad 1(b)$$

for positive electron affinity (PEA) ($\chi > 0$) and

$$QE = A(k_BT_e)^2[2k_BT_e + |\chi|] \qquad 2(a)$$

$$MTE = \frac{|\chi|}{2} + k_BT_e\left[\frac{3k_BT_e+|\chi|}{2k_BT_e+|\chi|}\right] \qquad 2(b)$$

for NEA ($\chi < 0$). In equations (1) and (2) $k_B$ is Boltzmann's constant, and $A$ is a constant that scales the QE due to various effects such as the details of the photoexcitation physics, carrier loss by recombination, etc. that are not included in the theoretical analysis. These expressions are shown below to well describe experimental measurements of the emission properties of studied Cesiated GaAs [44] and bare GaN(0001) photocathodes. The presented formalism can also be readily extended using numerical techniques to emitting bands with more complex dispersion characteristics and emission from multiple bands.

Of immediate note is that for large PEA the MTE tends to a limiting value of $k_BT_e$ (equation 1(b)) and the QE falls as $\exp[-\chi/k_BT_e]$ (equation 1(a)) as should be expected for emission from the tail of a thermalized Boltzmann electron distribution. Further, at $\chi = 0$, the MTE $= \frac{3}{2}k_BT_e$ and the QE is simply proportional to $(k_BT_e)^3$, implying that lowering the electron temperature has a stronger adverse effect on the emission efficiency. And, for NEA, the MTE cannot be reduced below $\frac{1}{2}|\chi|$ (equation 2(b)) – an offset simply associated with the energy-conserving mapping of the emitting electron distribution onto the vacuum states (Figure 1(b)). In addition, it is important to note that the initial momentum of the electron in the emitting band state is not a factor because an intermediate particle (the optical phonon) is involved in the momentum-resonant FC emission process.

Although the FC analysis employs phonon scattering with a zero optical phonon energy (i.e., $\hbar\Omega = 0$), which directly describes emission from the polaron (the quasi-particle generated by the coupling of conduction band electrons with ionic crystal vibrations [70,71]), it is readily modified for FC processes involving the emission of $n$ phonons through a redefinition of the electron affinity; specifically, $\chi \rightarrow \chi + n\hbar\Omega$. The strength of the $n^{th}$ FC emission process is given by Poisson statistics as $e^{-g}g^n/n!$ [72], where $g$ is the dimensionless Fröhlich (or electron-(optical)phonon) coupling constant associated with optical deformation scattering [73]. For GaAs, $g = 0.068$ [71], so that over 93% of the emission occurs for $n = 0$ (i.e., direct emission from the polaron). On the other hand, for GaN ($g \approx 0.4$ [74]), the $n = 0$, 1, 2, 3, and 4 emission modes are required to describe the FC emission process to an uncertainty of less than 0.1%. This means that an appropriately weighted sum is required to evaluate the total QE and hence also the MTE of electron emission. Fortunately, our analysis indicates that, to a very good approximation, the same results are obtained by evaluating equations (1) and (2) assuming that an average of $g$ optical phonons are involved in the FC emission process; that is, by using an effective electron affinity of $\chi + g\hbar\Omega$. We note that this energy shift, $g\hbar\Omega$, is also equal to the real part of the electron self-energy shift above the (conduction) band minimum associated with the polaron quasi-particle [61,70]. This means that the $\hbar\Omega = 0$ analysis presented in the Appendix (i.e., $\chi$ evaluated using the energetic minimum of the emitting electronic band (Figure 1)) is an accurate approximation for (optical)phonon-mediated FC electron emission from semiconductor photocathodes.

**Cesiated GaAs(001)**

Figure 2 displays the measured QE and MTE (black data points) for a reflection-mode Cesiated $p$-type GaAs photocathode illuminated by 808nm light – a photon energy ($\hbar\omega$ = 1.53eV) just above the 1.42eV band gap of the semiconductor where the absorption coefficient is $1.3\times10^4$cm$^{-1}$ [66]. The semiconductor photocathode consists of a 2.4μm GaAs active layer with $10^{19}$cm$^{-3}$ Zn doping atop a 0.3μm $Al_{0.55}Ga_{0.45}As$ buffer layer grown on a $n$-type GaAs(001) substrate. The procedure used to obtain the experimental data is detailed in Refs. 44 and 75. Briefly, the QE and MTE are measured using the transverse energy spread spectrometer (TESS) photocathode characterization system equipped with laser light sources [76,77]. After an initial activation using a Cs-O yo-yo deposition, which produced an evaluated electron affinity of around –0.1eV, the photocathode's performance was monitored at 300K as a function of the electron affinity under its controlled degradation (i.e., increase in electron affinity) through oxygen poisoning.

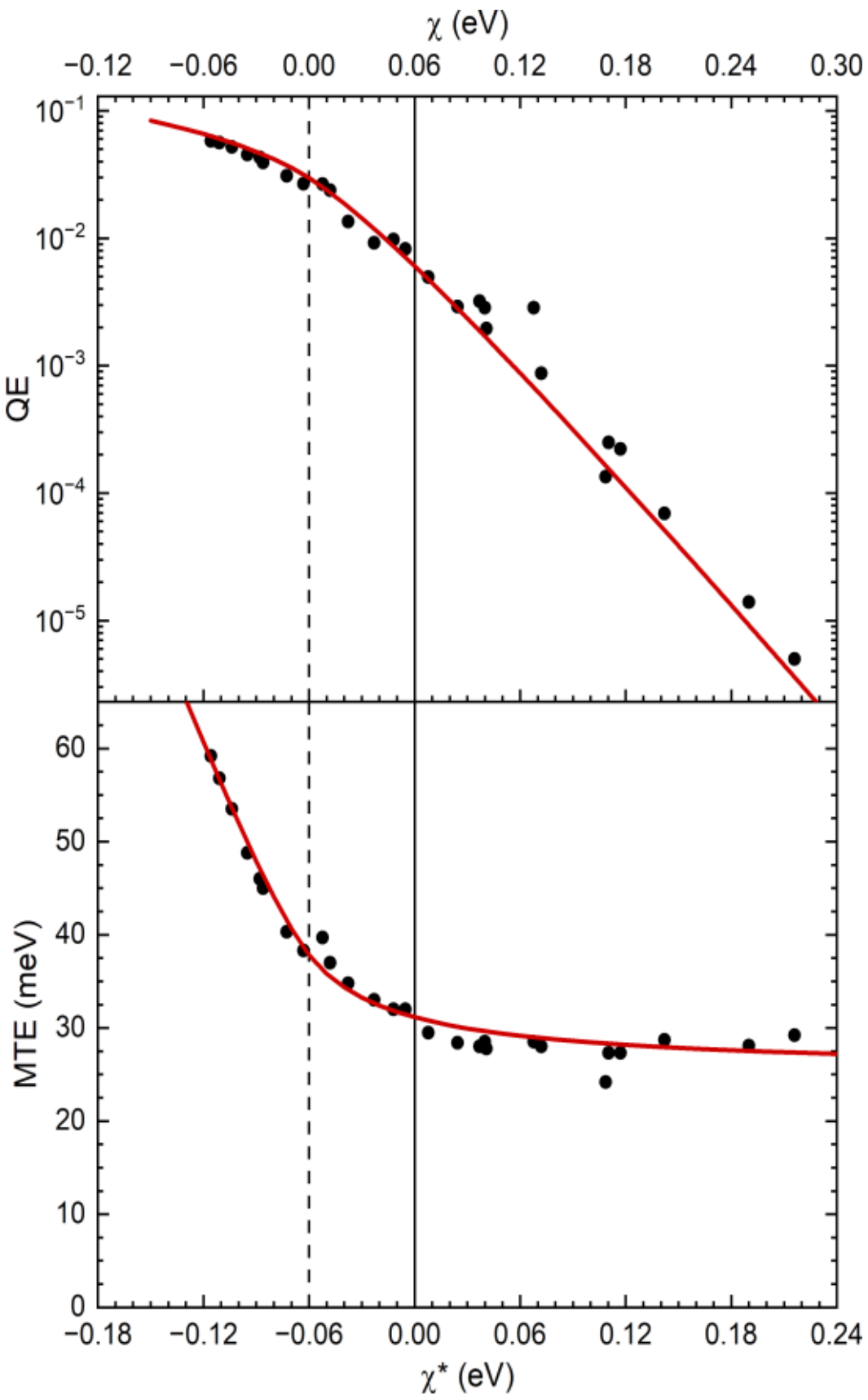

**Figure 2**
Experimental data (black points) and theoretical fits (red lines) for the QE (top panel) and MTE (bottom panel) as a function of the evaluated electron affinity $\chi^*$ (after Ref. 44) for a Cesiated $p$-type GaAs photocathode. The electron affinity $\chi = \chi^* + 60$meV determined using the measured [44] and theoretical (see Appendix) longitudinal electron energy distribution curves is shown on the top axis and $\chi = 0$ by the vertical dashed line.

The displayed optimum fit (red lines in Figure 2) of the theoretical QE and MTE dependencies (equations (1) and (2)) to the experimental data require a 60meV shift towards negative electron affinity; in other words, $\chi^* = \chi - 60$meV. This shift is due to the difference in the definition of the electron affinity which for the presented theoretical FC emission model is defined as shown in Figure 1; that is, the energy difference between the minimum of the emitting bulk band and zero energy in the vacuum. In contrast, the plotted experimental value of the electron affinity, $\chi^*$, was determined (with an estimated uncertainty of less than 13meV) by fitting the measured longitudinal emitted electron energy distribution to an expected functional form convolved with a 125meV FWHM

Gaussian instrument (resolution) function [44,75]. A direct comparison of the longitudinal energy distribution predicted by the FC emission mechanism (obtained by integration of the emitted electron distributions over the transverse momentum $p_T$, as described in the Appendix) and their measurement [44] indicates that the TESS instrument resolution is closer to 10meV rather than 125meV, and so the latter is the major contributing factor to the 60meV difference between $\chi^*$ and $\chi$. Both theoretical fits to the experimental QE and MTE data employ $k_BT_e$ = 27meV (i.e., an electron temperature just above 300K) – the limiting value of the MTE for large PEA (bottom panel of Figure 2 and equation 1(b)). This is to be expected since the long ~1μm absorption depth (i.e., transport length) in GaAs at 808nm provides the photoexcited electron distribution in the conduction band sufficient time for optical phonon emission to cool to near the lattice temperature while internal and applied fields accelerate the electrons towards the emission face. The analysis presented in Ref. 73 for this equilibrium case of electron drift with $T_e/T = 1.06$, where $T$ is the 300K lattice temperature, indicates $\mu_e E_{av} \approx 0.25 v_{ds}$, where $\mu_e$ is the electron mobility, $E_{av.}$ is the average acceleration field experienced by the electrons, and $v_{ds}$ is their saturation drift velocity. For GaAs (optical phonon energy $\hbar\Omega \approx$ 36meV and conduction band effective mass $m^* = 0.067m_0$), $v_{ds} \approx \sqrt{3\hbar\Omega/(4m^* \coth(\hbar\Omega/2T))}$ = $2.1\times10^5$m/s [73] and $\mu_e \approx$ 1,600cm$^2$/(V.s) for the $10^{19}$cm$^{-3}$ dopant density [79], implying that $E_{av.}$ ~ 3kV/cm. This value for the average acceleration field is somewhat short of the expected ~100MV/m surface depletion field of the $p$-type GaAs photocathode (assuming ~0.5eV band bending [79]) and so suggests that either (i) the Cesiation of the surface significantly reduces the band bending (through the introduction of new surface states) or (ii) the majority of the emitted electrons originate from bulk states beyond the ~10nm surface depletion region. The latter is quite likely since the transport analysis in Ref. 73 also indicates that the electrons drift with a velocity of $\mu_e E_{av.} \approx 0.25 v_{ds}$ = $5\times10^4$m/s and hence, including the acceleration due to the internal fields, the time for an electron to transit the depletion region is somewhat less than the characteristic ~200fs optical phonon emission time in GaAs [73,80].

**GaN(0001)**

In contrast to the GaAs photocathode, the emission surface of the studied $p$-type wurtzite GaN(0001) photocathode from Kyma Technologies Inc. [81] is not Cesiated. The photocathode is fabricated using hydride vapor-phase epitaxy (HVPE) techniques on a sapphire substrate and consists of a 1μm-thick 'active' layer with $5\times10^{17}$cm$^{-3}$ Mg doping grown atop a 2μm buffer layer of undoped GaN. The epi quality (sub-1nm rms surface roughness) GaN(0001) surface is N-polar terminated due to the growth orientation of the sapphire substrate [82,83] for which in-house DFT-based thin-slab calculations of the work function [84] indicate a value of 5.8(±0.2)eV as measured from the valence band maximum – a value that is consistent with prior work [85]. For the 3.4eV band gap of GaN at 300K [86], this implies that the conduction band minimum (CBM) has a PEA of about 2.4eV which will severely limit emission of near room temperature electrons as the Boltzmann factor $\exp[-E/k_BT] \sim 10^{-41}$. On the other hand, the band structure of wurtzite GaN [87] also indicates the presence of an upper conduction band ~2.5eV above the CBM which, if populated, would allow more efficient and likely NEA electron emission.

Surface band bending due to the $p$-type Mg doping of the GaN semiconductor generates a depletion region with a ~40nm depth (assuming 1.0-1.8eV of band bending [88]) which is of the same order as the 50-100nm absorption depth for incident photon energies just above the 3.4eV band gap [86]. For the resultant average depletion field of ~10MV/m, the room temperature electron drift velocity is 1-2$\times10^5$m/s [86,89-91] (and likely higher for hot photoexcited electrons), giving a transit time over the depletion region significantly less than 1ps; that is, less than the expected ~1ps optical phonon decay time at our $5\times10^{17}$cm$^{-3}$ dopant density [92]. This means that electrons photoexcited into the conduction band with an excess energy $\Delta E$ (with respect to the CBM) will initially (on a sub-100fs timescale within the ~0.5ps incident laser pulse duration) equipartition their energy with the $l = 6$ primary Γ-point optical phonon modes in GaN [93] and this equipartitioned energy should be maintained during electron emission as they drift to the GaN(0001) photocathode face. The electron temperature $T_e$ of the emitting electron distribution for an incident photon energy $\hbar\omega$ is then given by

$$k_BT_e = \frac{2\Delta E}{3(l+1)} = \frac{2\mu(\hbar\omega - E_g)}{3m_e(l+1)} \approx \frac{1}{12}\left(\hbar\omega - E_g\right) \tag{3}$$

in the parabolic band approximation, where $m_e$ = $0.22m_0$ and $m_h \approx 1.25m_0$ are the effective masses of the conduction band [86,89] and valence band (heavy hole) [95] states near the Γ point, respectively, and the reduced mass $\mu = m_e m_h/(m_e + m_h)$.

Figure 3 displays the measured spectral dependence of the MTE and QE for the $p$-type GaN(0001) photocathode at 300K. The experimental data was obtained using the sub-picosecond tunable UV laser and the 10-20kV DC gun-based photocathode characterization system described in Refs. 95 and 96. As expected, the QE (top panel of Figure 3) shows an onset at around the 3.4eV band gap energy and is overall much smaller than that measured for the Cesiated GaAs photocathode (Figure 2) due to the ~2.4eV PEA of the CBM. Also noticeable is a non-zero QE below the band gap energy which we ascribe to photoexcitation into the conduction band from the Mg dopant states [29] 100-200meV above the valence band maximum [97]. These low excess energy photoexcited electrons are generated throughout the 1μm-thick *p*-doped GaN region due to the low (less than $10^4$cm$^{-1}$) absorption coefficient in this spectral region. This means that below $\hbar\omega \approx 3.4$eV the photoexcited electrons will drift into the surface depletion region at close to room temperature (as was the case for the GaAs photocathode irradiated at 808nm), but then undergo acceleration in the dominant internal field of the ~40nm surface depletion region. For the average ~10MV/m depletion field strength and the expected ~400cm$^2$/V.s electron mobility at the $5\times10^{17}$cm$^{-3}$ dopant density [98], the transport analysis in Ref. 73 indicates that an electron temperature close to the 880K Debye temperature $T_D$ of GaN [99] should be attained (i.e., $k_BT_e \approx k_BT_D = 75$meV), leading to the sub-bandgap MTE signals in Figure 3. We note here that the ~3nm product of the characteristic ~20fs optical phonon scattering time in GaN [73,100] with the 1-2×$10^5$m/s electron drift velocity [86,89-91] is much less than the ~40nm width of the surface depletion region, thus ensuring that a thermalized electron distribution is maintained.

For incident photon energies greater than the $E_g \approx$ 3.4eV band gap, band-to-band absorption dominates driving the photocathode's electron emission dynamics into the short transport length regime where $k_BT_e$ is dependent upon $\hbar\omega - E_g$ (equation 3). Indeed, the red line in the bottom panel of Figure 3 employs $k_BT_e = k_BT_D + \frac{1}{12}(\hbar\omega - E_g)$ for $E_g > 3.3$eV and $\chi \approx -0.3$eV to fit the experimental MTE data using equation (2b). The latter negative value for the electron affinity, which is required to describe the 230meV offset in the measured MTE at $\hbar\omega \approx E_g$, confirms that the observed emission originates from electrons in the upper conduction band 2.5eV above the CBM in GaN. For the downward ~1eV band bending [88] and our calculated intrinsic $\chi = -0.1(\pm0.2)$eV for the upper conduction band, this value of an 'effective' electron affinity for the FC emission process suggests that only electrons well within the photocathode's ~40nm surface depletion region are emitted through the optical phonon scattering mechanism. In fact, assuming parabolic band bending towards the *p*-type GaN(0001) surface, a 0.2eV band energy increase (about one fifth of the band bending) is achieved ~5nm from the surface – a distance corresponding to ~10 lattice constants that may reflect the spatial overlap required in GaN between the bulk and emitted vacuum electron wavefunctions plus that of the scattering optical phonon (i.e., lattice vibration) for the FC emission process.

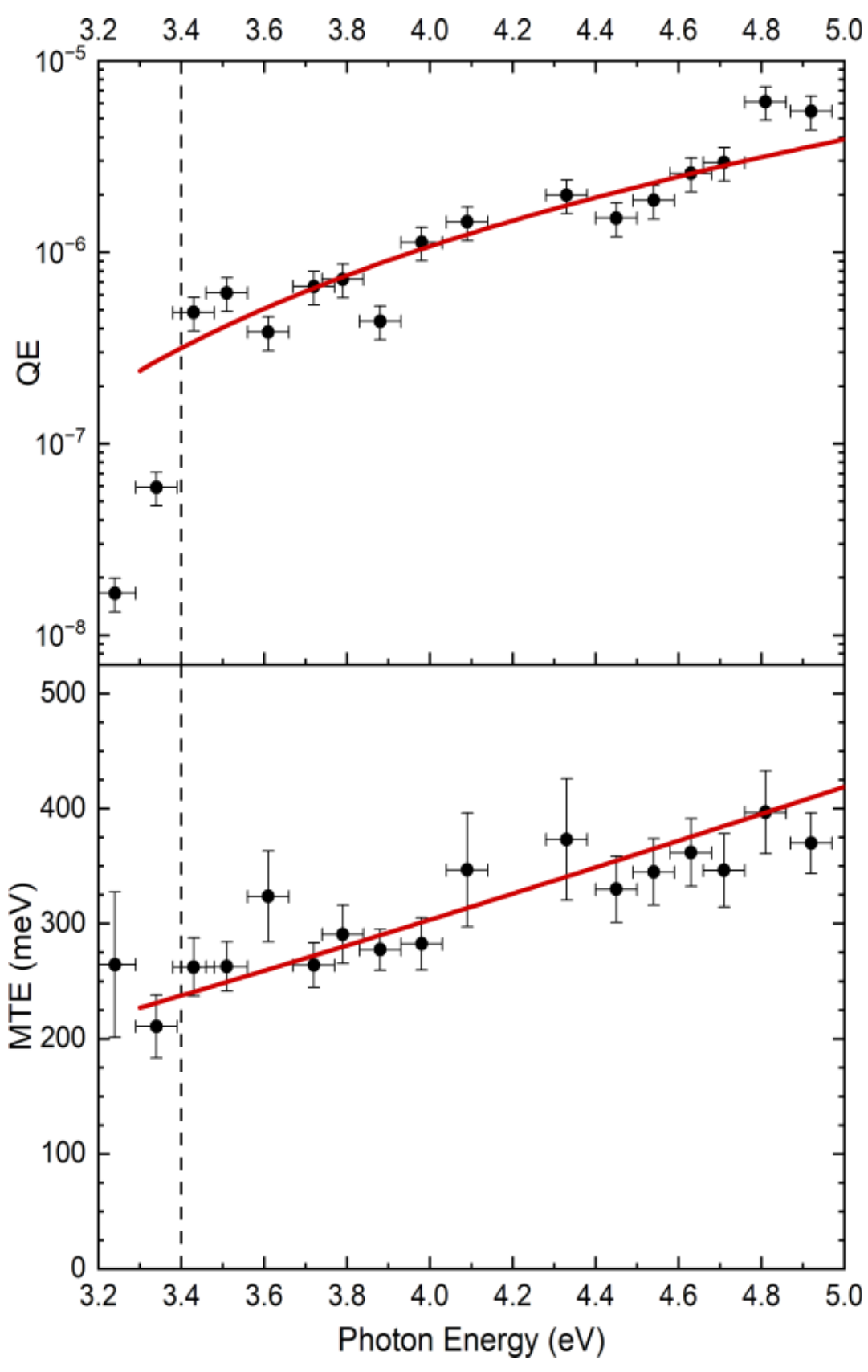

**Figure 3**
Experimental data (black points) and theoretical fits (red lines) for the QE (top panel) and MTE (bottom panel) as a function of the incident photon energy for the *p*-type GaN(0001) photocathode. The 3.4eV bandgap of GaN is indicated by the vertical black dashed lines.

Using equation (2a), the measured spectral variation of the QE is also consistent with a NEA of 0.3eV and the same $k_BT_e$ dependence on $\hbar\omega$ used to fit the MTE data as illustrated by the red line fit in the top panel of Figure 3. Further, the ~$10^{-6}$ QE observed over most of the investigated spectral range is quite consistent with the average 100-150meV electron energy $k_BT_e$ due to the $\exp[-E/k_BT_e] \sim 10^{-8}$ Boltzmann factor and the estimated factor of 100 difference in the density of states between the upper and lower conduction bands.

**Summary**

An analytical theoretical formulation of (optical)phonon-mediated FC electron emission for polar semiconductor photocathodes is presented that accurately describes the measured spectral emission properties of a *p*-type GaN(0001) photocathode and those at 808nm of a Cesiated *p*-type GaAs photocathode as its electron affinity is degraded (i.e., increased) by controlled oxidation at 300K [44]. Specifically, experimental QE and MTE data obtained for both photocathodes are fit using the FC emission theory presented in the Appendix with electron temperatures of $k_BT_e$ = 27meV for GaAs and $k_BT_e = k_BT_D + \frac{1}{12}(\hbar\omega - E_g)$ when ħω > 3.3eV for GaN – the values expected for their long and short transport length regimes respectively. Further, the surface roughness of the two photocathodes is somewhat different; the (0001) GaN surface is of epi polished quality with a sub-1nm rms surface roughness whereas the (001) GaAs photocathode surface is somewhat rougher due to the Cesiation and subsequent oxidation with a rms surface roughness of 2-6nm being reported [101-103]. Consequently, the clear implication is that sub-10nm rms surface roughness does not play a major role in electron emission from polar semiconductors – a conclusion consistent with a recent multi-dimensional analysis of the effects of surface roughness [104]. Further, a FC electron emission process should not be strongly affected by surface roughness effects since it is energy and momentum resonant; that is, an electron can only be emitted into the vacuum if the scattering process in the solid-state photocathode produces an electronic state that satisfies $E = p^2/(2m_0)$ with a positive *z*-component (longitudinal) momentum for any polar (θ) emission angle.

The presented momentum-resonant FC process is expected to dominate the electron emission of all solid-state photocathodes with sufficiently strong optical deformation potential scattering; the Fröhlich coupling constant *g* for GaAs is only 0.068 [71], whereas that for both diamond (*g* = 1.1 [61]) and GaN (*g* ≈ 0.4 [74]) are much larger. Highly polar semiconductor photocathode materials like $Cs_2Te$, $Cs_3Sb$, and the bi-alkali antimonides should certainly exhibit dominant electron emission due to the (optical)phonon-mediated FC mechanism. We note, for example, that the limiting (near threshold) value of the 300K MTE measured for $Cs_3Sb$ at 690nm (ħω = 1.797eV) is 35-40meV [20] rather than 25meV. Recent *ab initio* band structure calculations for cubic $Cs_3Sb$ [105] indicate that this could be expected for $k_BT_e$ ≈ 25meV (the long transport length limit): An upper conduction band exists near the vacuum level at 0.4-0.5eV above its CBM giving χ ≈ 0 for emission from this band (as $Cs_3Sb$ has a PEA of around 0.45eV [18]) so that equations 1(b) and 2(b) for the FC emission mechanism predict $MTE \approx \frac{3}{2}k_BT_e$ = 38meV. Notably, this 300K MTE measurement for *p*-type $Cs_3Sb$ has also been attributed, using Monte Carlo modeling techniques, to photoexcitation into the conduction band from excess Sb acceptor states ~0.4eV above the valence band and subsequent electron emission that does not conserve transverse momentum [106]. Clearly, the former is energetically equivalent to the upper conduction band from band structure calculations while the latter is a partial physical representation of the FC emission process presented here.

**Acknowledgements**

This work was supported by the U.S. Department of Energy under Award no. DE-SC0020387 and part-funded by UKRI-STFC through the ASTeC core grant. The authors are also indebted to H.E. Scheibler for sharing access to experimental data from Ref. 44 and clarifying the analysis methods employed.

**Appendix**

Within the parabolic band approximation, an analytical expression for momentum-resonant Franck-Condon (FC) electron emission mediated by optical phonon scattering can be derived. For the case of a step potential photoemission barrier, the over-barrier transmitted electron flux $\mathbf{j}_z$ in the longitudinal (*z*) direction perpendicular to the photocathode surface may be expressed as

$$j_z = \frac{p_{0z}}{m_0}|T|^2 = \frac{p_z}{m^*}(1-|R|^2) = \frac{4p_{0z}p_z^2}{m^*(p_{0z}+p_z)^2}$$

where the emission is from a bulk band with an energy dispersion $E = \frac{p^2}{2m^*}$ to a vacuum band with $E_0 = \frac{p_0^2}{2m_0}$ ($m^*$ is the effective mass of the electron-like bulk band, $m_0$ is the free electron mass, *p* is the electron momentum in the emitting band, and $p_0$ is the emitted electron momentum in the vacuum), and $|T|^2$ and $|R|^2$ are the usual transmission and reflection probabilities for the square barrier. For momentum-resonant emission, $p_z = p_{0z}$, so that the transmitted electron flux simply becomes

$$j_z = \frac{p_{0z}}{m^*} = \frac{p_0 cos\theta}{m^*}$$

where θ is the polar emission angle. This angle and momentum dependence for the transmitted electron flux is incorporated in the following analysis of FC electron emission for the simplified case where the energy of the mediating (optical)phonon ħΩ is taken to be zero; that is, much smaller than the characteristic energy of the electron distribution $k_B T_e$, where $k_B$ is Boltzmann's constant and $T_e$ is the electron temperature. As a result, the mathematical description of the emission process becomes a straightforward energy-conserving mapping of the electron distribution in the emitting bulk band states that are above the vacuum level onto the vacuum density of states (DOS), as any vacuum state may be populated through the momentum-resonant FC emission mechanism. Positive ($\chi > 0$) and negative ($\chi < 0$) electron affinity cases for the emitting band are considered separately and the derived expressions for the MTE and QE collapse to the same value for zero electron affinity, i.e., $\chi = 0$.

***i) Positive Electron Affinity***

For the case of positive electron affinity (PEA), where the minimum of the emitting band is below the vacuum level, so that $E = E_0 + \chi$ , the momentum-space energy mapping requires

$$p^2 = \left(\frac{m^*}{m_0}\right) p_0^2 + 2m^*\chi$$

which allows a three-dimensional non-degenerate electron distribution from $E$ to $E + dE$ in the emitting bulk band, namely $\sqrt{E}.exp[-E/(k_B T_e)]$ , to be written in the vacuum momentum space as $\left\{\left(\frac{m^*}{m_0}\right) p_0^2 + 2m^*\chi\right\} exp\left[-\frac{p_0^2}{2m_0 k_B T_e} - \frac{\chi}{k_B T_e}\right]$ for the interval $p_0$ to $p_0 + dp_0$ . Integration in spherical co-ordinates over the positive $p_{z0}$ momentum half-space in the vacuum DOS with the flux transmission dependence then gives the number of emitted electrons $N$;

$$N = 2\pi A \int_0^{\frac{\pi}{2}} d\theta sin\theta cos\theta \int_0^{\infty} dp_0\, p_0^2 \left(\frac{p_0}{m^*}\right)\left\{\left(\frac{m^*}{m_0}\right) p_0^2 + 2m^*\chi\right\} exp\left[-\frac{p_0^2}{2m_0 k_B T_e} - \frac{\chi}{k_B T_e}\right]$$

$$= \pi A\, (2m_0 k_B T_e)^2 [2k_B T_e + \chi]\, exp\left[-\frac{\chi}{k_B T_e}\right]$$

where the integration over the azimuthal angle ϕ has been performed and $A$ is a constant containing all effects not associated with the emission physics: dimensional constants for the vacuum DOS, spin degeneracy, carrier recombination, etc. The above expression therefore provides the expected dependence of the barrier emission QE on $T_e$ and $\chi$ for the PEA FC process; in other words

$$QE \propto (k_B T_e)^2 [2k_B T_e + \chi]\, exp\left[-\frac{\chi}{k_B T_e}\right]$$

The MTE of the emitted electrons can now be evaluated using the suitably normalized expectation value $\langle\frac{p_{0T}^2}{2m_0}\rangle = \langle\frac{p_0^2 sin^2\theta}{2m_0}\rangle$ , where $p_T$ is the transverse electron momentum, viz;

$$MTE = \frac{\pi A}{N m^* m_0} \int_0^{\frac{\pi}{2}} d\theta sin^3\theta cos\theta \int_0^{\infty} dp_0\, p_0^5 \left\{\left(\frac{m^*}{m_0}\right) p_0^2 + 2m^*\chi\right\} exp\left[-\frac{p_0^2}{2m_0 k_B T_e} - \frac{\chi}{k_B T_e}\right]$$

$$\Rightarrow\ MTE = k_B T_e \left[\frac{3k_B T_e + \chi}{2k_B T_e + \chi}\right]$$

The integral above for $N$ may also be evaluated in cylindrical co-ordinates and thus separately over the transverse and longitudinal vacuum momenta, respectively, to give emitted electron distributions for PEA as a function of their longitudinal ($E_z$) and transverse ($E_T$) vacuum energies:

$$N(E_z) = C k_B T_e \sqrt{E_z}\, \{E_z + \chi + k_B T_e\}\, exp\left[-\frac{(E_z + \chi)}{k_B T_e}\right]$$

and

$$N(E_T) = C k_B T_e\, \{E_T + \chi + k_B T_e\}\, exp\left[-\frac{(E_T + \chi)}{k_B T_e}\right]$$

where $C$ is a constant.

***ii) Negative Electron Affinity***

For a negative electron affinity (NEA) photocathode, the minimum of the emitting band is above the vacuum level, implying that $E = E_0 - |\chi|$, so that the momentum-space energy mapping is

$$p^2 = \left(\frac{m^*}{m_0}\right) p_0^2 - 2m^*|\chi|\, .$$

Hence, the three-dimensional non-degenerate electron distribution in the emitting bulk band can be written in vacuum momentum space as $\left\{\left(\frac{m^*}{m_0}\right) p_0^2 - 2m^*|\chi|\right\} exp\left[-\frac{p_0^2}{2m_0 k_B T_e} + \frac{|\chi|}{k_B T_e}\right]$ for the interval $p_0$ to $p_0 + dp_0$ and $p_0 \geq \sqrt{2m_0|\chi|}$ . The number of emitted electrons is then given by

$$N = 2\pi A \int_0^{\frac{\pi}{2}} d\theta sin\theta cos\theta \int_{\sqrt{2m_0\chi}}^{\infty} dp_0\, p_0^2 \left(\frac{p_0}{m^*}\right)\left\{\left(\frac{m^*}{m_0}\right) p_0^2 - 2m^*|\chi|\right\} exp\left[-\frac{p_0^2}{2m_0 k_B T_e} + \frac{|\chi|}{k_B T_e}\right]$$

$$= \pi A\,(2m_0 k_B T_e)^2[2k_B T_e + |\chi|]$$

As a result, in terms of $T_e$ and $\chi$, the barrier emission QE of the NEA FC process is of the form

$$QE \propto (k_B T_e)^2[2k_B T_e + |\chi|]$$

Similarly, the MTE is again readily evaluated as for the PEA case using

$$MTE = \frac{\pi A}{N m^* m_0}\int_0^{\frac{\pi}{2}} d\theta sin^3\theta cos\theta \times \int_{\sqrt{2m_0|\chi|}}^{\infty} dp_0\, p_0^5 \left\{\left(\frac{m^*}{m_0}\right) p_0^2 - 2m^*|\chi|\right\} exp\left[-\frac{p_0^2}{2m_0 k_B T_e} + \frac{|\chi|}{k_B T_e}\right]$$

$$\Rightarrow\ MTE = \frac{|\chi|}{2} + k_B T_e\left[\frac{3k_B T_e + |\chi|}{2k_B T_e + |\chi|}\right]$$

Clearly, as $\chi \to 0$, the evaluated MTEs for both the PEA and NEA cases equal $\frac{3}{2}k_B T_e$ and the QE values also converge to the same value.

Again, the integral above for $N$ may be evaluated in cylindrical vacuum momentum co-ordinates to give emitted electron distributions for NEA as a function of their longitudinal ($E_z$) and transverse ($E_T$) vacuum energies:

$$N(E_z \le |\chi|) = C(k_B T_e)^2\sqrt{E_z}$$

$$N(E_z > |\chi|) = Ck_B T_e\sqrt{E_z}\,\{E_z - |\chi| + k_B T_e\}\, exp\left[-\frac{(E_z - |\chi|)}{k_B T_e}\right]$$

and

$$N(E_T \le |\chi|) = C(k_B T_e)^2$$

$$N(E_T > |\chi|) = Ck_B T_e\,\{E_T - |\chi| + k_B T_e\}\, exp\left[-\frac{(E_T - |\chi|)}{k_B T_e}\right]$$

where $C$ is a constant.

---